\documentclass[10pt, letterpaper, conference]{IEEEtran}
\usepackage{amsthm}
\usepackage{cite}
\usepackage{multirow}
\usepackage{tabularx}
\usepackage[hyphens]{url}

\ifCLASSINFOpdf
\usepackage[pdftex]{graphicx}
\else
\usepackage[dvips]{graphicx}
\fi
\graphicspath{{figs/}}

\theoremstyle{definition}
\newtheorem{definition}{Definition}
\IEEEoverridecommandlockouts

\begin{document}
\title{Regional Consistency: Programmability and Performance for
Non-Cache-Coherent Systems}
\author{\IEEEauthorblockN{Bharath Ramesh, Calvin J. Ribbens}
\IEEEauthorblockA{Center for High-End Computing Systems\\
Department of Computer Science, Virginia Tech, Blacksburg, VA 24061\\
Email: \{bramesh,ribbens\}@cs.vt.edu}
\and
\IEEEauthorblockN{Srinidhi Varadarajan}
\IEEEauthorblockA{Dell Inc.\\
Email: \{Srinidhi\_Varadarajan@dell.com\}}}
\IEEEaftertitletext{\vspace{-2\baselineskip}}
\maketitle

\begin{abstract}
Parallel programmers face the often irreconcilable goals of programmability and
performance. HPC systems use distributed memory for scalability, thereby
sacrificing the programmability advantages of shared memory programming models.
Furthermore, the rapid adoption of heterogeneous architectures, often with
non-cache-coherent memory systems, has further increased the challenge of
supporting  shared memory programming models. Our primary objective is to define
a memory consistency model that presents the familiar thread-based shared memory
programming model, but allows good application performance on non-cache-coherent
systems, including distributed memory clusters and accelerator-based systems. We
propose \emph{regional consistency} (RegC), a new consistency model that
achieves this objective. Results on up to 256 processors for representative
benchmarks demonstrate the potential of RegC in the context of our prototype
distributed shared memory system.
\vspace{-1\baselineskip}
\end{abstract}

%\begin{IEEEkeywords}
%memory consistency; relaxed consistency; cache coherence; distributed
%shared memory; parallel programming;
%%\vspace{-\baselineskip}
%\end{IEEEkeywords}

\section{Introduction}
A fundamental issue on all high performance computing platforms is how best to
share data among concurrent tasks. For example, parallel applications may share
data between physically distributed nodes of a cluster, between processor and
coprocessor or accelerator on a single node, among multiple coprocessors
attached to a single node, between cores of a single processor or accelerator,
and among various components of a cluster-on-a-chip. Each of these scenarios
presents unique challenges and opportunities to the system designer; but each
has one important feature in common, namely that there is a distinction between
local and remote memory. Application developers use local ``cached'' data to
gain performance by exploiting spatial and temporal locality.

From a programmer's point of view, the key question is what programming model
should be used to orchestrate concurrency and data sharing across these
memories. The most straightforward model is probably the traditional shared
memory model, e.g., as offered by POSIX Threads (Pthreads) over cache-coherent
shared-memory hardware. On such platforms data sharing is transparent, and
simple synchronization mechanisms allow programmers to write correct and
performant codes. This traditional shared memory model is dominant for
platforms with the largest market share, e.g., portable devices, laptops,
servers. Hence, there is a growing ecosystem of shared memory parallel programs,
tools and design practices.

Unfortunately, two dominant trends in high-end computing---scalable clusters and
heterogeneous nodes---work against the traditional shared memory model.
Distributed memory clusters offer no physically shared memory at all. The
dominant programming model on clusters is explicit message-passing. Other
proposed models and systems for clusters come closer to the traditional shared
memory model, including Partitioned Global Address Space (PGAS)~\cite{pgas}
languages and distributed shared memory (DSM) systems~\cite{l88,fp89,zsm90,
cbz91,bzs93,kcdz94,sgt96,m97,dhkns99,hst99,ri00,np04}. Meanwhile, emerging
heterogeneous node architectures generally do offer shared memory. However, they
do not provide cache coherence across all components of the system, and the best
programming model for these systems is still an open question.

Resolving this tension---between a programmer's desire for a strong shared
memory consistency model and an architect's need to sacrifice cache-coherence
for scalability and heterogeneity---requires a new look at memory consistency
models. The consistency model defines the semantics of memory accesses; to a
large extent it determines both the performance and programmability of the
programming model. In this paper we propose \emph{regional consistency} (RegC),
a new memory consistency model that gives programmers the strong shared memory
programming model they prefer, but that can be implemented efficiently over
modern non-cache-coherent systems.

\begin{table*}
\caption{Comparison of properties of popular consistency model.}
\label{tab:models}
\centering
\begin{tabular}{|lccll|}
\hline
\multicolumn{1}{|c}{\multirow{2}{*}{Consistency model}} &
Shared data and synchronization & Separates critical and &
\multicolumn{1}{c}{\multirow{2}{*}{Consistency granularity}} &
\multicolumn{1}{c|}{non-critical section} \\
& primitive association & Non-critical section accesses & &
\multicolumn{1}{c|}{accesses data consistency}\\
\hline
Entry consistency & explicit & yes & object & no consistency guarantees\\
Scope consistency & transparent & yes & page & requires barriers \\
Release consistency & transparent & no & page & consistent\\
\multirow{2}{*}{Regional consistency} & \multirow{2}{*}{transparent} &
\multirow{2}{*}{yes} & page - ordinary region & \multirow{2}{*}{consistent} \\
& & & object - consistency region & \\
\hline
\end{tabular}
\vspace{-\baselineskip}
\end{table*}

The common approach to providing shared memory semantics over non-cache-coherent
architectures is to relax the consistency model to allow greater parallelism in
data access, but at the cost of some ease of programming. Table~\ref{tab:models}
compares three popular relaxed consistency models with RegC in terms of four
defining properties: 1) whether shared data must be explicitly associated with
synchronization primitives; 2) whether critical and non-critical section memory
accesses are distinguished; 3) the granularity at which consistency updates are
typically done; and 4) to what extent consistency is maintained for non-critical
memory updates. The entry consistency model~\cite{bzs93} requires explicit
association between shared data and synchronization primitives; it does not
provide any memory consistency guarantees for non-critical-section data.
Although entry consistency has performance advantages, the association
restriction and its lack of consistency for updates to shared data done outside
of critical sections make it difficult to use. Scope consistency~\cite{isl96}
removes the explicit association requirement and makes non-critical section
updates consistent at barriers. Scope consistency unburdens the programmer from
the explicit shared data to synchronization primitive association but
consistency updates are done at the granularity of a page, which has performance
implications. Release consistency~\cite{gllggh90} does not require any explicit
association nor does the programmer have to use barriers to make non-critical
section data consistent. In this sense, release consistency is easier to program
than either entry or scope consistency. However, since release consistency does
updates at the granularity of a page, and since it does consistency updates at
page granularity for both critical sections and non-critical sections it can
still suffer from performance problems.

Our new RegC memory consistency model is motivated by these observations. It
explicitly distinguishes between modifications to memory protected by
synchronization primitives and those that are not, allowing for a more
performant and scalable implementation. In this paper we give performance
results for two implementations of RegC---one that takes full advantage of this
distinction and one that does not---in order to emphasize and evaluate the
relative benefits of this feature. In our first implementation of RegC all
updates are done at the granularity of a page. In the more sophisticated
implementation we use fine grain (object level) updates for modifications to
shared data protected by synchronization primitives and use page invalidations
for modifications to data not protected by synchronization primitives. In
essence, RegC is similar to entry consistency for critical-section shared data
accesses and similar to release consistency for non-critical section accesses.

Our current implementations of RegC are part of \emph{Samhita}, a portable
user-level distributed shared memory (DSM) system. Samhita/RegC provides
cache-coherent shared memory semantics over the physically distributed memory of
a cluster. Compared to other platforms where such a programming model is desired
but not directly supported by the hardware (e.g., processor+coprocessor,
cluster-on-a-chip), a distributed memory cluster is in some sense the hardest
case, since there is no shared memory at all, and network latency and bandwidth
can be a significant bottleneck. However, we believe the rapid and steady
improvements in high-end interconnect performance (relative to memory latency
and bandwidth) allow us to treat DSM primarily as ``just'' another cache
management problem, and as an excellent testbed for evaluating the potential of
the RegC memory consistency model.

The architecture of Samhita and early performance results are described
in~\cite{rrv11}. The focus of this paper is on regional consistency. We give
performance results that evaluate Samhita's implementation of RegC, and also
identify how the runtime system can support additional performance-enhancing
extensions. The remainder of the paper is organized as follows. In
Section~\ref{sec:rw} we present an overview of related work on memory
consistency models. We define and describe the regional consistency model in
Section~\ref{sec:regc}. A brief overview of Samhita and its implementation
appears in Section~\ref{sec:samhita}.  Performance results from one
computational kernel and two applications are presented and discussed in
Section~\ref{sec:perf}. We conclude and discuss future work in
Section~\ref{sec:cfw}.

\section{Related work}
\label{sec:rw}
For a programmer to write correct concurrent applications, the results of memory
operations need to be predictable. Memory consistency models describe the rules
that guarantee memory accesses will be predictable. There are several memory
consistency models that have been proposed, including sequential consistency
(SC)~\cite{l79}, weak consistency (WC)~\cite{dsb86}, processor consistency
(PC)~\cite{g89}, release consistency (RC)~\cite{gllggh90}, entry consistency
(EC)~\cite{bzs93}, scope consistency (ScC)~\cite{isl96}.

Sequential consistency (SC) has two important properties: (1) program order is
maintained at each processor, (2) global order is an interleaving of all the
sequential orders at each processor. The SC model, though conceptually simple,
is extremely strong and imposes restrictions that negatively affect performance.
To alleviate the performance limitations of SC other consistency models have
been proposed that relax or weaken the restrictions.

The weak consistency (WC) model, one of the earliest weak models, differentiates
shared data into two categories: data that has no effect on concurrent
execution, and data that includes synchronization variables to protect access to
shared data or provide synchronization. WC has three main characteristics:(1)
access to all synchronization variables is sequentially consistent, (2) no
operation on synchronization variables is permitted until all previous accesses
to shared data are performed, and (3) no access to shared data is allowed until
all previous operations on a synchronization variable have been performed. An
important distinction between WC and SC is that consistency is enforced on a set
of accesses for WC rather than individual accesses. Weak consistency improves
performance by overlapping writes from a single processor; to serialize writes
to the same location programmers are required to use synchronization variables,
which is an added burden. Processor consistency (PC) follows a middle approach
between WC and SC. PC allows writes from two processors, observed by themselves
or a third processor not to be identical. However, writes from any processor are
observed sequentially.

One of the biggest drawbacks in weak consistency is the fact that when a
synchronization variable is accessed, the processor has no knowledge if the
access to the shared data is complete or about to start. This requires the
processor to perform a memory consistency operation every time a synchronization
variable is accessed. Release consistency (RC) extends WC by categorizing
accesses to shared data as \emph{ordinary} or \emph{special} accesses, which are
equivalent to accesses to data and synchronization variables, respectively. RC
further categorizes \emph{special} accesses as \emph{sync} and \emph{nsync}
accesses. Finally, \emph{sync} access are further categorized as either
\emph{release} or \emph{acquire} accesses (analogous to the corresponding mutex
lock operations). The RC model enforces the following rules: (1) before any
ordinary read or write access is performed, all previous \emph{acquire} accesses
must be performed, (2) before a \emph{release} access is performed all previous
reads and writes done by the processors must be performed, and (3) all accesses
to synchronization variables are processor consistent. At every release the
processor propagates its modifications to shared data to all other processors.
This entails a significant data transfer overhead. To reduce the amount of data
transfer, propagation of modified data is postponed in a variant known as lazy
release consistency (LRC)~\cite{kcz92}. In LRC the acquiring processor
determines the modification it requires to meet the requirements of RC.

Both the WC and RC models use synchronization primitives to ensure ordering of
access to shared data. Entry consistency (EC) exploits this relationship between
synchronization primitives and access to shared data by requiring all shared
data to be explicitly associated with at least one synchronization primitive.
Whenever a synchronization primitive is acquired all updates to the shared data
associated with that synchronization primitive are performed. In EC each
synchronization primitive has a current owner that last acquired the primitive.
When the ownership changes because another processor acquires the
synchronization primitive, all updates to the shared data associated with the
primitive are sent to the acquiring processor. To reduce performance impact,
synchronization primitives can exist in two modes---\emph{exclusive} and
\emph{non-exclusive}. In the \emph{non-exclusive} mode, though the
synchronization primitive is owned by one processor it can be replicated at
others. Only a single processor is allowed to acquire a synchronization
primitive in \emph{exclusive} mode. To modify the shared data associated with a
synchronization primitive a processor must own the synchronization primitive in
\emph{exclusive} mode.

Though association of shared data with synchronization primitives reduces the
overhead of data transfer among processors, EC is hindered by the increased
complexity of explicitly associating shared data with synchronization
primitives. Programming using EC is complicated and can be error prone. Scope
consistency (ScC) alleviates the explicit association of shared data with
synchronization primitives. ScC detects the association dynamically at the
granularity of pages, thus providing a simpler programming model. The implicit
association of memory accesses to synchronization primitives is termed the
\emph{consistency scope}. The ScC model defines the following rules: (1) before
a new session of a consistency scope is allowed to be open at a processor, all
previous writes performed with respect to the scope need to be performed at the
processor and (2) access to memory is allowed at a processor only after all the
associated consistency scopes have been successfully opened. Though ScC presents
a relaxed consistency model, the programming model exposed to the user is
complex when compared to RC or LRC. Iftode et al.~\cite{isl96} mention that
precautions need to be taken to ensure that a program runs correctly under ScC,
the primary challenge being that all accesses to shared data must be made inside
critical sections.

All of the previously discussed consistency model are sequentially consistent
for data race free codes. The authors of location consistency (LC)~\cite{gs00}
present a model that is not sequentially consistent, i.e., writes to the same
location are not serialized and not necessarily observed in the same order by
any processor. LC represents the state of a memory location as a partially
ordered multiset of write and synchronization operations. For the LC model to be
able to provide this partial ordering of writes and synchronization operations
it requires an accompanying cache consistency model which is not provided by
traditional multi-processor systems. Because writes to the same location are not
serialized the programming model associated with using LC is complicated and
adds a significant burden on the programmer.

To summarize, programmability and performance are two ends of a spectrum. The
traditional approach in the past to enable performance on parallel platforms was
to use a relaxed consistency model. However, weaker consistency models achieve
performance by sacrificing programmability. In our approach, to support the
familiar memory consistency model expected by today's shared memory programmers,
we provide a strong consistency model. However, we believe most of the
performance can be recovered by a consistency model that enables one to develop
intelligent runtime system that support it and by providing programmers with
extensions to the programming model that can leverage intrinsic information
available only at runtime.

\section{Regional consistency}
\label{sec:regc}
\begin{figure}
	\centering
	\includegraphics[scale=0.80]{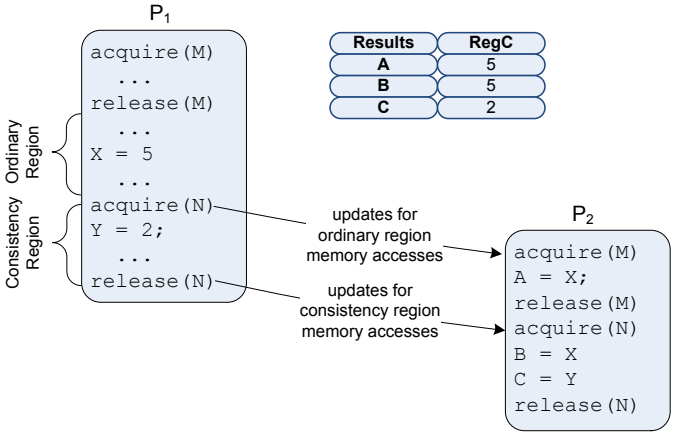}
	\caption{Pseudo code describing regional consistency.}
	\label{fig:regc}
	\vspace{-1\baselineskip}
\end{figure}

Before giving a formal definition of our new consistency model, we describe the
basic idea and how it compares to similar models. The idea behind \emph{regional
consistency} (RegC) is to divide an application's memory accesses into two kinds
of regions---\emph{consistency regions} and \emph{ordinary regions}---as
depicted in Figure~\ref{fig:regc}. These regions are demarcated by
synchronization primitives utilizing mutual exclusion (mutex) locks and
barriers. More specifically, a consistency region is demarcated by a mutex lock
acquire and release. All memory accesses outside of a consistency region occur
in an ordinary region. A barrier separates one ordinary region from another,
i.e., one ordinary region ends at a barrier and a new one begins after a
barrier.

The RegC rule for barriers is simple: all modifications made in the preceding
ordinary region are made consistent for the processors participating in that
barrier. To describe the RegC rules for consistency regions, we first define a
\emph{span} as one instance of a consistency region that executes at a given
processor. A span starts at the acquire of a mutex lock and ends on the
successful release of that lock. Any modification to data made in a span will
be visible to processors that subsequently enter spans corresponding to the
same mutex lock. Note that spans corresponding to different locks are
independent, i.e., they can execute concurrently. Different spans can also be
nested, corresponding to nested critical sections. Finally, modifications made
in the preceding ordinary region are propagated on the start of a span. RegC
guarantees that these updates will be visible at other processors before the
start of \emph{any} span corresponding to \emph{any} consistency region.

Regional consistency can be viewed as an amalgamation of release consistency and
scope consistency. Similar to ScC, we transparently detect data modification
within a consistency region and implicitly associate it with corresponding
locks, thereby creating the dichotomy of ordinary and consistency accesses.
Similar to RC, we ensure that updates from ordinary regions are propagated on
lock acquisition/release, not just on explicit barrier operations. We believe
that performing updates from ordinary regions only on explicit barriers is
unduly restrictive, i.e., it limits parallel problem decomposition to block
synchronous codes. For other common parallel decompositions (e.g.,
producer/consumer, pipeline) superimposing barrier semantics creates unnecessary
synchronization between unrelated threads and increases false sharing.

The general view is that relaxing consistency models improves performance but at
the cost of programmability. Since our goal with RegC is to maintain the
familiarity of the strong consistency model expected by thread-based programs,
the challenge is to allow for a performant implementation of the consistency
model. Both RegC and RC provide a sufficiently strong model for writing correct
threaded code compared to ScC. The differences between RegC and RC allow
significant performance opportunities for RegC. Explicitly distinguishing
between memory modifications made inside a critical section and those made
outside allows an implementation of RegC to delay updates made in ordinary
regions, which RC cannot (LRC, which makes a similar optimization, is less
intuitive to programmers than RegC.) Furthermore, the distinction allows a RegC
implementation to use different update policies to propagate the modifications
in ordinary and consistency regions, i.e., page-based invalidation policy for
ordinary regions and fine grained updates for consistency regions.

\subsection{Formalizing RegC}
To define RegC formally we use the formal definitions for the memory access
transitions presented in~\cite{sd87}. For the purpose of completeness we include
these definitions here:

\begin{definition}{\bf{Performing with respect to a processor}.}
	A $LOAD$ by processor $P_i$ is considered \emph{performed} with respect to
	$P_k$ at a point in time when the issuing of a $STORE$ to the same address by
	$P_k$ cannot affect the value returned by the $LOAD$. A $STORE$ by $P_i$ is
	considered \emph{performed} with respect to $P_k$ at a point in time when an
	issued $LOAD$ to the same address by $P_k$ returns the valued defined by this
	$STORE$ (or a subsequent $STORE$ to the same location).
\end{definition}

\begin{definition}{\bf{Performing an access globally}.}
%An access is \emph{performed} when it is \emph{performed} with respect to all
%processors.
	A $STORE$ is \emph{globally performed} when it is \emph{performed} with
	respect to all processors. A $LOAD$ is \emph{globally performed} if it is
	\emph{performed} with respect to all processors and if the $STORE$ that is the
	source of the returned value has been \emph{globally performed}.
\end{definition}

In addition to the above two standard definitions we propose the following new
definition.

\begin{definition}{\bf{Subsequently after}.}
	A \emph{span} for any \emph{consistency region} at $P_j$ is said to start
	\emph{subsequently after} a \emph{span} for any \emph{consistency region} at
	$P_i$ if and only if the \emph{span} has successfully started at $P_i$ before
	the \emph{span} at $P_j$ successfully starts. Note that a span only
	successfully starts when the corresponding lock acquisition succeeds.
\end{definition}

Before we define the RegC model formally, we distinguish a $STORE$
\emph{performed} with respect to the regions of memory accesses as follows:
\begin{itemize}
	\item A $STORE$ \emph{performed} within a \emph{consistency region} is defined
		as a \emph{consistent} $STORE$.
	%e.g., a $STORE$ that occurs within a
	% critical section protected by a lock.
	\item A $STORE$ \emph{performed} outside of a \emph{consistency region} is
		defined as an \emph{ordinary} $STORE$.
	%e.g., a $STORE$ that occurs outside
	%of a critical section protected by a lock.
\end{itemize}

Furthermore, we distinguish a \emph{consistent} $STORE$ being \emph{performed}
with respect to a \emph{consistency region} from a $STORE$ being
\emph{performed} with respect to a processor as follows:
\begin{itemize}
	\item A \emph{consistent} $STORE$ is \emph{performed} with respect to a
		\emph{consistency region} when the current \emph{span} of that
		\emph{consistency region} ends.
	\item A $STORE$ is \emph{performed} with respect to $P_i$ if a subsequent
		$LOAD$ issued by $P_i$ returns the value defined by this $STORE$ (or a
		subsequent $STORE$ to the same memory location).
\end{itemize}

The rules for \emph{regional consistency} are as follows:
\begin{enumerate}
	%\item An \emph{ordinary} $STORE$ \emph{performed} at $P_i$ must be
	%\emph{performed} with respect to $P_j$ after the successful start of a
	%\emph{span} of any \emph{consistency region} at $P_j$, if and only if the
	%\emph{span} starts \emph{subsequently after} the \emph{span} of the
	%\emph{consistency region} that started successfully at $P_i$ after the
	%\emph{ordinary} $STORE$ was \emph{performed} at $P_i$.

	\item Before a span is allowed to start on $P_j$ subsequently after a span on
		$P_i$, any ordinary $STORE$ performed at $P_i$ before that span on $P_i$
		must be performed with respect to $P_j$.

	\item Before a new \emph{span} of a \emph{consistency region} is allowed to
		successfully start at $P_i$, any \emph{consistent} $STORE$ previously
		\emph{performed} with respect to that \emph{consistency region} must be
		\emph{performed} with respect to $P_i$.

	\item A $STORE$ \emph{performed} at $P_i$ must be \emph{performed} with
		respect to $P_j$, for all $P_i$ and $P_j$ participating in a barrier.
\end{enumerate}

The first rule determines when an \emph{ordinary} $STORE$ is performed with
respect to a processor. An \emph{ordinary} $STORE$ is performed with respect to
a processor $P_j$ before any \emph{span} is allowed to start at $P_j$, provided
this \emph{span} starts \emph{subsequently after} the \emph{span} immediately
following the $STORE$ at processor $P_i$. The second rule ensures that when a
new \emph{span} starts at a processor, any \emph{consistent} $STORE$
\emph{performed} previously with respect to that \emph{consistency region} is
guaranteed to be \emph{performed} with respect to the processor. The third rule
guarantees that any $STORE$ performed before the start of a barrier is performed
with respect to all processes participating in that barrier.

\section{Overview of Samhita}
\label{sec:samhita}
Samhita solves the problem of providing a shared global address space by casting
it as a cache management problem. This motivates our approach of separating the
notion of serving memory from the notion of consuming memory for computation.
Each Samhita thread is associated with a local cache; the entire shared global
address space is accessed through this local cache. This cache can be considered
another level in the memory hierarchy. Efficient cache management can hide the
latency difference between accessing local memory and remote memory.

Samhita's architecture consists of compute servers, memory servers and resource
managers. These three components execute on the physical nodes of a cluster. The
compute servers execute one or more threads of control from one or more
applications. Samhita exposes a {\it fork-join} execution model similar to POSIX
threads. It is important to note that individual threads of an application
correspond to traditional processes in Samhita. The runtime system
transparently shares the global data segment of a program across different
processes. Because of this transparent sharing, processor cores on physically
different nodes appear as individual cores of a shared memory system. The memory
servers combine to provide the global shared address space. To mitigate the
impact of hot spots, memory allocations are strided across multiple memory
servers. The total size of the global address space is equal to the combined
amount of memory exported by the memory servers. The resource manager is
responsible for job startup, thread placement, memory allocation, and
synchronization.

To highlight and evaluate the benefits of distinguishing between ordinary and
consistency regions, we consider two implementations of RegC in Samhita.
In the first version, though we distinguish between ordinary and consistency
region stores, there are still performance limitations
due to the consistency granularity being a page for both ordinary and
consistency region updates. For example, if only a small amount of data on a
given page is updated in a consistency region, we must still invalidate the
entire page on the corresponding lock acquisition in any thread. Using fine
grain updates on lock acquire, similar to entry consistency, is a better
approach in this scenario and is implemented in our second version. When using
fine grain updates we need to propagate only the changes made to individual
shared variables in a consistency region. However, this requires us to track
individual stores. We do this by instrumenting all the stores an application
performs using the LLVM compiler framework~\cite{la04}. We insert a function
call to the runtime system before each store is performed. The runtime tracks
stores performed within a consistency region, and ignores those performed in an
ordinary region. On lock release we are then able to propagate the changes made
in the consistency region. Our experimental evaluation suggests that the
overhead incurred by such store instrumentation is within reasonable limits for
most applications. We believe that we can further reduce the store
instrumentation overhead by not instrumenting most ordinary stores using static
analysis on the application code. This is part of our future work.

\section{Performance evaluation}
\label{sec:perf}
In this section we describe performance studies that demonstrate that Samhita
and RegC provide a programmable, scalable, and efficient shared memory
programming model. We compare our two different implementations of RegC to
underline the performance benefits achieved by using fine grain updates for
consistency regions and page based invalidation for ordinary regions. In the
rest of this section \emph{samhita} refers to the implementation that uses fine
grain updates for consistency regions and page invalidation for ordinary region
updates, and \emph{samhita\_page} refers to the implementation that uses page
invalidation for both consistency and ordinary region updates.

We present scalability results on up to 256 cores, which to our knowledge is the
largest scale test by a significant margin for any DSM system reported to date.
The results demonstrate that for scalable algorithms, our Samhita implementation
achieves good weak scaling on large core counts. Strong scaling results for
Samhita are very similar to equivalent Pthreads implementations on a single
node. For less scalable algorithms, scalability is limited by synchronization
overhead. We identify extensions to the programming model that transparently
leverage information about data placement and consistency requirements to
improve performance.

The performance evaluation was carried out on {\it System~G}, a 2600 core
cluster (325 nodes). Each node is a dual quad-core 2.8GHz Intel Xeon (Penryn
Harpertown) with 8GB of main memory. The cluster is interconnected over a quad
data rate (QDR) Infiniband switched fabric. The first set of results are for a
synthetic benchmark consisting of our thread based implementation of the STREAM
TRIAD~\cite{m95}. We then present results from two application benchmarks:
Jacobi and molecular dynamics applications based on codes from the
OmpSCR~\cite{drsg05} repository. We ported the OmpSCR benchmarks from OpenMP to
the equivalent threaded code. In fact, to emphasize the similarity between the
Pthreads API and the Samhita API our two implementations for each benchmark are
derived from the same code base. Memory allocation, synchronization and thread
management calls are represented by macros, which are processed using the {\tt
m4} macro processor.

The performance evaluation includes strong scaling results where problem size is
fixed as the number of cores grows, and weak scaling experiments where problem
size grows with the number of cores. The strong scaling experiments use
a single memory server. The weak scaling experiments use 20 memory servers to
accommodate the largest problem size.

\subsection{STREAM TRIAD}
\begin{figure}
	\centering
	\includegraphics[scale=0.65]{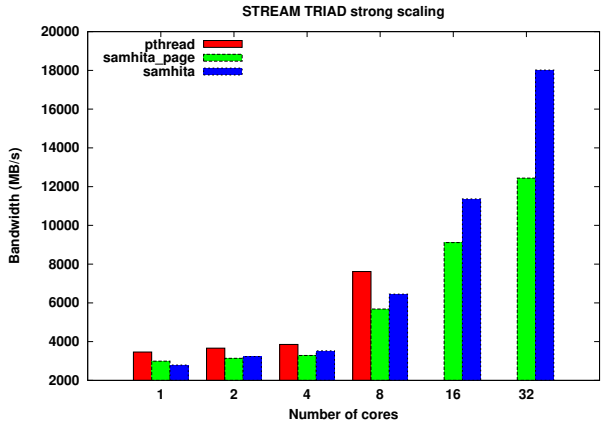}
	\caption{Sustained memory bandwidth vs.\ number of cores for STREAM TRIAD
	synthetic benchmark. Vectors of dimension $n=16M$.}
	\label{fig:st_s}
	%\vspace{-\baselineskip}
\end{figure}

\begin{figure}
	\centering
	\includegraphics[scale=0.65]{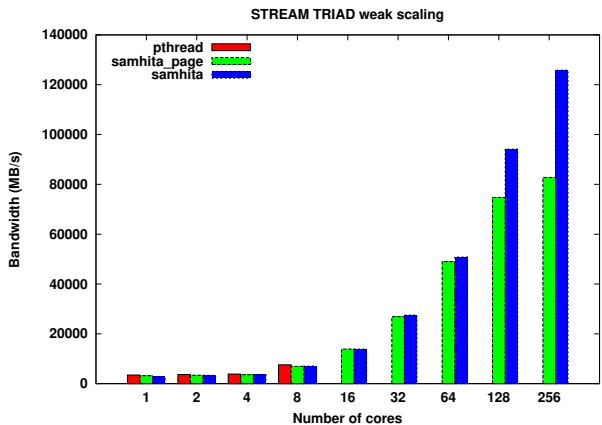}
	\caption{Sustained memory bandwidth vs.\ number of cores for Samhita and
		Pthreads based STREAM TRIAD benchmark. Data size $3n$ and number of cores
		$p$ are scaled proportionally, i.e., $3n/p$ is a constant. Problem size for
		$p=1$ is $n=16M$; problem size for $p=256$ is $n=4G$.}
	\label{fig:st_w_ic}
	\vspace{-\baselineskip}
\end{figure}

\begin{figure}
	\centering
	\includegraphics[scale=0.65]{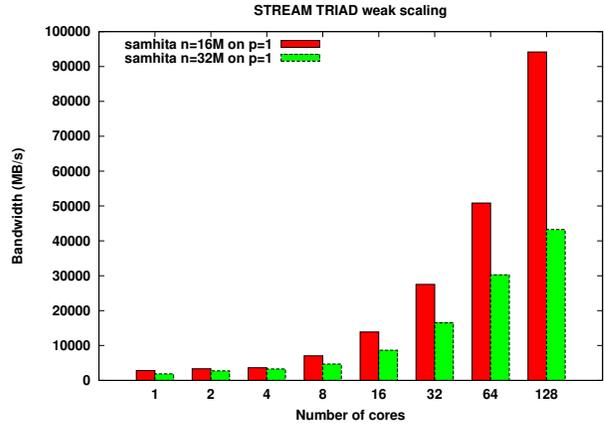}
	\caption{Sustained memory bandwidth vs.\ number of cores for Samhita for two
		different problem sizes. Data size $3n$ and number of cores $p$ are scaled
		proportionally, i.e., $3n/p$ is a constant. Problem size for small problem
		on $p=1$ node is $n=16M$; problem size for large problem on $p=1$ node is
		$n=32M$.}
	\label{fig:st_w_oc}
	\vspace{-\baselineskip}
\end{figure}

The STREAM benchmark is a synthetic benchmark that measures sustained memory
bandwidth for a set of simple vector kernels. We implemented a thread based
version of the TRIAD operation. This operation is a simple vector update (or
DAXPY), a level 1 operation from the BLAS package. The TRIAD kernel computes $A
= B + \alpha C$, where $A$, $B$ and $C$ are vectors of dimension $n$ and
$\alpha$ is a scalar. Each run of the benchmark consists of 400 iterations of
the TRIAD operation with a barrier between each iteration.

Figure~\ref{fig:st_s} compares the strong scaling bandwidth achieved by the
Pthreads and the two Samhita implementations. The Samhita implementations
achieve a reasonable sustained bandwidth, which scales as we increase the number
of cores. The bandwidth achieved by the \emph{samhita} implementation is close
to $85\%$ of that achieved by the Pthreads implementation for the 8 core run,
while \emph{samhita\_page} achieves $74\%$ of the memory bandwidth. We note that
the bandwidth achieved for $1$--$4$ cores is similar due to the fact that our
physical nodes are dual socket and our placement policy fills the first socket
before filling the second socket.

Figure~\ref{fig:st_w_ic} presents weak scaling results for up to 256 processors.
The performance of both Samhita implementations tracks Pthreads up to 8 cores
and continues to scale well up to 128 cores for \emph{samhita\_page} and 256
cores for \emph{samhita}, before synchronization costs begin to constrain
scalability.

In the weak scaling results shown in Figure~\ref{fig:st_w_ic} the data
associated with each process fits entirely in the Samhita cache associated with
that process. Figure~\ref{fig:st_w_oc} shows the same results for
\emph{samhita}, along with results for a problem size twice as big. The larger
problem no longer fits in the local Samhita cache, which results in capacity
misses; the entire data must be streamed in for each iteration. We see that when
the resulting data spills occur there is a clear impact on the achieved
bandwidth. However, we also notice that the Samhita implementation still
continues to scale reasonably well; we lose at most a factor of two despite
having to refill the cache on each iteration with data served from remote memory
servers. This illustrates the benefit of our optimization for fetching remote
pages and the benefits of the simple prefetching strategy used in our current
implementation.

\subsection{Jacobi}
\begin{figure}
	\centering
	\includegraphics[scale=0.65]{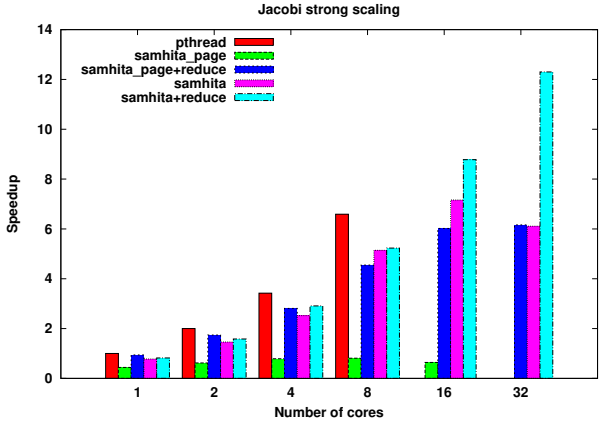}
	\caption{Parallel speedup vs.\ number of cores for Jacobi. Speedup is
		relative to 1-core Pthreads execution time.}
	\label{fig:jacobi_s}
\end{figure}

\begin{figure}
	\centering
	\includegraphics[scale=0.65]{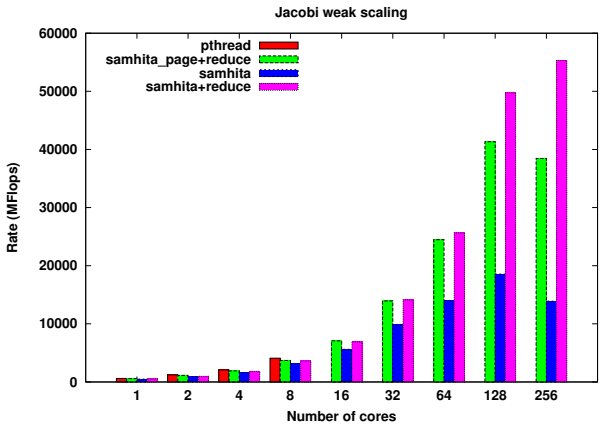}
	\caption{Computation rate vs.\ number of cores for Samhita and Pthreads based
		Jacobi. Data size $3n^2$ and number of cores $p$ are scaled proportionally,
		i.e., $3n^2/p$ is a constant. Problem size for $p=1$ is $n=4096$; problem
		size for $p=256$ is $n=65536$.}
	\label{fig:jacobi_w}
	\vspace{-\baselineskip}
\end{figure}

The Jacobi benchmark application is a threaded implementation of the Jacobi
OpenMP code found in OmpSCR~\cite{drsg05}. It corresponds to an iterative linear
solver applied to a finite difference discretization of a Helmholtz equation.
Figure~\ref{fig:jacobi_s} compares the strong scaling speedup of Pthreads and
four Samhita implementations of the Jacobi benchmark. The Pthreads
implementation and two Samhita implementation use a mutex variable to protect
the global variable that accumulates the residual error on each iteration.
Barrier synchronizations are required at three points of each iteration as well.
We notice that the lock-based Samhita implementation does not speed up as the
number of processors increases for the \emph{samhita\_page} implementation. The
reason for the performance degradation is the strong memory consistency provided
by RegC. Performance profiling shows that the majority of the time is spent in
one of barriers, which follows the consistency region, and requires expensive
memory consistency operations to reflect the memory updates made in the
preceding ordinary region. The \emph{samhita} lock based implementation on the
other hand shows good strong scaling results up to 16 processors. This
improvement in performance can be solely attributed to the fine grain updates to
propagate the changes made in a consistency region.

Relaxing the consistency model would improve performance, but programmability
would be sacrificed. Instead, we extend the programming model by providing a
reduction operation (as in OpenMP) that replaces the operation performed in the
consistency region but is implemented by the Samhita runtime system. The second
set of Samhita results in Figure~\ref{fig:jacobi_s} show the dramatic
improvement in performance. Using the reduction operation extension, the
\emph{samhita\_page} implementation achieves just over $69\%$ of the speedup
achieved by Pthreads for the 8-core run. Using the reduction extension in the
\emph{samhita} case also yields performance improvement achieving just over
$79\%$ of the speedup achieved by Pthreads, but the improvement is not as
dramatic as in the \emph{samhita\_page} case. This once again underlines the
importance of having different update mechanisms for ordinary and consistency
regions.

Figure~\ref{fig:jacobi_w} presents weak scaling results for Jacobi on up to 256
processors. We see that all the Samhita implementations of Jacobi track the
Pthreads case very well up to 8 cores. The benefit of the reduction extension is
again clear, with good scalability for both Samhita implementations to 128
cores.  Beyond 128 cores the scalability of this algorithm is limited, i.e., as
problem size and core counts grow, the cost of synchronization eventually
outweighs the computation.

\subsection{Molecular dynamics}
\begin{figure}
	\centering
	\includegraphics[scale=0.65]{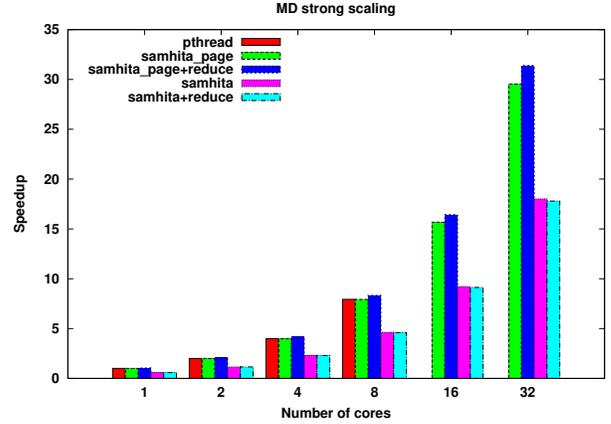}
	\caption{Parallel speedup vs.\ number of cores for molecular dynamics. Speedup
		is relative to 1-core Pthreads execution time.}
	\label{fig:md_s}
	\vspace{-\baselineskip}
\end{figure}

The molecular dynamics application benchmark is a simple n-body simulation using
the velocity Verlet time integration method. The particles interact with a
central pair potential. The OpenMP code from OmpSCR uses reduction operations
for summing the kinetic and potential energy of the particles. Similar to the
Jacobi benchmark, our threaded implementations use a mutex variable to protect
the global kinetic and potential energy variables. Barrier synchronization is
used during various stages of the computation for synchronization.
Figure~\ref{fig:md_s} compares the strong scaling speedup of Pthreads and the
Samhita implementations.

We see that the two different Samhita implementations, one using mutex variables
and the other using reduction variables, track the Pthreads implementation very
closely for the \emph{samhita\_page} implementation.  For the \emph{samhita}
case, we notice that though the application scales well there is a visible
impact of the cost associated with the store instrumentation. In this
application the cost associated with synchronization is significantly lower than
the computational cost. Most of the stores are performed in ordinary regions but
the instrumentation function is still called. We can use static analysis of the
application code to avoid instrumenting most ordinary region stores. We believe
that with this approach we can reclaim most of the lost performance due to
instrumentation overhead. This benchmark result clearly indicates that
applications that are computationally intensive (the computation per particle is
$O(n)$) can easily mask the synchronization overhead of Samhita enabling the
application to scale very well.

\section{Conclusions}
\label{sec:cfw}
We have defined \emph{regional consistency} (RegC), a new memory consistency
model that allows programs written using familiar threading models such as
Pthreads to be easily ported to a non-cache-coherent system. We evaluated the
performance of two implementations of RegC using Samhita, a system that provides
shared memory over a distributed memory cluster supercomputer. Recent advances
in modern high performance interconnects allow us to implement a relatively
strong consistency model (easier programmability) while still achieving
acceptable application performance using a sophisticated runtime system.

Performance results show that our Samhita implementations achieve computational
speedup comparable to the original Pthreads implementations on a single node
with trivial code modifications, and illustrate the performance improvements
achieved by a simple programming model extension and by distinguishing ordinary
and consitency region stores. Weak scaling results on up to 256 processor cores
demonstrate that scalable problems and algorithms scale well over Samhita.

A promising future enhancement is be to use static analysis to avoid
instrumenting most ordinary region stores, thus reclaiming most of the
performance overhead associated with store instrumentation in our current
implementation. Another possible direction for future research is exploring the
additional performance improvement opportunity by delaying propagation of
updates done in ordinary region, similar to lazy release consistency
(LRC)~\cite{kcz92}. We also plan to investigate providing a shared memory
programming model using regional consistency by extending Samhita to other
non-cache-coherent platforms like accelerators, cluster-on-chip and
coprocessors.

\bibliographystyle{IEEEtran}
\bibliography{ccgrid2013}
\end{document}